# Deformability-based circulating tumor cell separation with conical-shaped microfilters: concept, optimization and design criteria

Mohammad Aghaamoo,[1] Zhifeng Zhang,[1] Xiaolin Chen,[1] and Jie Xu [2,*]

[1] *Department of Mechanical Engineering, Washington State University, Vancouver, WA 98686 USA*

[2] *Department of Mechanical & Industrial Engineering, University of Illinois at Chicago, Chicago, Illinois 60607 USA*

[*] Corresponding author contact: jiexu@uic.edu, Phone: (312) 355-1788

Fax: (312) 413-0447

Circulating tumor cells (CTCs) separation technology has made positive impacts on the cancer science in many aspects. The ability of detecting and separating CTCs can play a key role in early cancer detection and treatment. In recent years, there has been growing interest in using deformability-based CTC separation microfilters due to their simplicity and low cost. Most of previous studies in this area are mainly based on experimental work. Although experimental research provides useful insights in designing CTC separation devices, there is still a lack of design guidelines based on fundamental understandings of the cell separation process in the filers. While experimental efforts face challenges especially microfabrication difficulties, we adopt numerical simulation here to study conical-shaped microfilters using deformability difference between CTCs and blood cells for separation process. We use liquid drop model for modeling a CTC passing through such microfilters. The accuracy of the model in predicting the pressure signature of the system is validated by comparing with previous experiments. Pressure-deformability analysis of the cell going through the channel is then carried out in detail in order to better understand how a CTC behaves throughout the filtration process. Different system design criteria such as system throughput and unclogging of the system are discussed. Specifically, pressure behavior under different system throughput is analyzed. Regarding the unclogging issue, we define pressure ratio as a key parameter representing the ability to overcome clogging in such CTC separation devices and investigate the effect of conical angle on the optimum pressure ratio. Finally, the effect of unclogging applied pressure on the system performance is examined. Our study provides detailed understandings of the cell separation process and its characteristics, which can be used for developing more efficient CTC separation devices.

## I. INTRODUCTION

Cancer is one of the most pervasive diseases and a major health concern all over the world. Each year, a lot of people die from cancer due to its high fatality ratio.[1] According to American Cancer Society, one in four deaths in this country is as a result of cancer[2] and this rapid growth will rise by 50% in the next few years due to population aging.[3] Cancer metastasis is recognized as the main reason of death due to cancer.[4] Circulating tumor cells (CTCs) play a key role in cancer metastasis; these cells are released from the primary tumor from a specific organ into the blood stream and spread through the cardiovascular system to different organs in the body and form a distant secondary tumor. Since CTCs have the same observable properties and genetic structure of the primary tumor cells and also play a key role for cancer metastasis, there is growing interest in developing efficient CTC separation techniques.[5, 6] There are technical challenges facing CTC separation devices such as extremely low percentage of CTCs in patient blood sample, different behavior of cancer cells and etc. Thus, more detailed studies of the cell process in such devices are required before designing and fabricating useful devices.

In general there are two broad CTC separation methods: biochemical methods and biophysical methods.[6] Biochemical methods mostly rely on biological markers or antibodies for recognizing and separating CTCs. One of the key challenges in these methods is the heterogeneity and genetic instability of tumor cells which make it difficult or in some cases unlikely to find a unique markers for the whole cells.[7] On the other hand biophysical approaches utilize physical markers such as cell size, shape, deformability, density and etc. to discriminate CTCs from blood cells.[8] These methods have the advantageous of label-free sorting. Recently, there have been growing interests in using microfluidic technologies to devise label-free CTC separation devices. By using this technology, higher capture efficiency and isolation purity can be reached because it enables us to precisely control the device parameter at cellular scale.[9] In addition, microfluidics has a great advantage of operating in laminar flow regime which provides more precise control on the cell manipulation.[10] Reduced cost and enhanced portability are among other advantages of this technology.[8]

Among various microfluidic CTC separation techniques, deformability-based CTC separation method has begun to emerge recently. This method has the advantages of simplicity and its potentially low-cost.[11] Moreover, deformability is a more efficient physical marker with respect to size since it has a stronger connection to cell phenotype.[12] Weir,[13] pillar,[14] pore[15] and channel[16] are among different types of CTC separation devices that are based on deformability as their main or one of their major physical markers. Most of the previous studies on deformability-based CTC separation devices are mainly based on experimental data and analysis. McFaul et al.[14] proposed a microstructural ratchet mechanism for cell separation process.

They utilized both size and deformability as separation physical marker. Non-uniformity of their proposed ratchet mechanism overcomes the challenges in unclogging conventional microfilters. Tan et al.[17] presented a label-free separation microdevice consists of arrays of crescent-shaped isolation wells as constrictions to distinguish CTCs from blood cells using their physical differences in deformability and size. Their proposed device offers high separation efficiency and ability of obtaining viable isolated cells. Tang et al.[18] introduced a microfluidic device with integrated microfilter of conical-shaped holes. Based on their studies, they suggested a (cell) nuclei size dependent design of microholes. In order to improve system performance, system parameters such as microfilter geometry and operating conditions need to be studied, understood and optimized. Numerical methods are powerful tools in optimization processes both in saving time and preventing the high cost of experimental research. Indeed, the aforementioned experimental work all relied on highly sophisticated microfabrication technologies. Currently, it is still quite challenging to fabricate arbitrary 3D geometries at microscale with high precision. Therefore, numerical methods are the only practical means of study at this point, with the hope that in near future more efficient experimental tools will become available and cheap enough for experimental implementation, such as the emergence of 3D printing technology in recent years.[19]

In a typical CTC separation process, passing event of a single CTC through the microfilter is a fundamental step which requires in-depth understanding. The key challenge in numerical simulation of a CTC filtering process is the mechanical modeling of living cells. In general, there are two main approaches in modeling living cells: micro/nanostructural approaches[20-24] and continuum approaches. In micro/nanostructural approaches the cell behaviors and responses are studied at the level of cytoskeleton (CSK) while in continuum approaches the cell is considered as a material with specific continuum material properties; this feature gives the continuum approach the advantages of being easier and more straightforward over the former approaches. It should be taken into account that the continuum approach is mostly suitable for the cases in which biomechanical response at the cell level is required. Generally, continuum model can be classified into two main categories: liquid drop models and solid models.[25] Based on this classification there are various research on modeling of the cellular entry process into a constriction channel. Yeung et al.[26] developed a cortical shell-liquid core model for passive flow of a cell into a tube at constant suction pressure. Leong et al.[27] numerically modeled breast cancer cell entering into a constricted micro-channel using compound liquid drop model. In their study they did some parametric studies on cell velocity profile, stiffness, elasticity and deformation profile in the process of cell passing through the constriction. Zhang et al. [28] simulated a CTC passing event through a microfiltering channel with constant cross sectional area based on liquid drop model. Their main goal was to characterize pressure signature of different types of cells passing through different channels over a wide range of flow rate. Luo et al. [29] used a cellular viscoelastic model (solid model) for modeling cellular entry process into a constriction channel using a finite element package ABAQUS. One of their main aims was to quantify instantaneous Young's modulus of the living cells based on their proposed model. Generally solid models are applicable for simulating small cellular deformation while for large cellular deformation it is better to use a liquid drop model.[27]

In the present research, we study the concept and design criteria of conical-shaped CTC separation microfilters as an example of non-uniform cross-sectional microfilters. The geometrical asymmetry design of such microfilters outperforms designs with symmetric geometries, in that it allows using periodically reversed flow for unclogging such filters without undoing the separation process.[14] Numerical simulation is utilized in order to better understand the basic principles behind deformability-based CTC separation devices and investigate the effect of system parameters on different design criteria of the system. Specifically, we model the process as a single CTC passing event through the microfilter. The simplification of just studying a single CTC is mainly based on the fact that the blood cells are much softer than CTCs and as a result they can easily pass the microfilter.[16] Thus, the main determinant factor in these devices is how a CTC can be squeezed through the channel. Since the main parameter in this process is pressure signature of the event, we use liquid drop model for mechanical simulation of CTC. Although this model greatly simplify the living cell, it can perfectly predict the pressure signature of the squeezing process as validated by Guo et al.[30] in their study on cell deformation in a micro-constriction. In the present study, numerical method is first validated with previously reported experiments. Pressure-deformability analysis of a CTC passing through the microfilter is then performed in order to better understand the concept of filtration process. Some important system design criteria of the system are introduced and the effects of system parameters on them are investigated. Specifically, the effect of system throughput on the critical pressure of the system is studied. We also discuss the clogging issues of deformability-based microfilters and investigate the effect of geometry of the microfilter and system applied pressure on solving system clogging.

## II. MODEL DESCRIPTION

The device filtering process is simply based on the flow of sample blood cells through a filtering channel in a microfluidic chip. There is a critical pressure defined for each cell to squeeze through the microfilter which strongly depends on cell surface tension coefficient, cell diameter and geometry of the microfilter. The more deformable and smaller cells can simply pass the channel while the stiffer and larger ones are required to overcome large critical pressure and will be blocked if there is not enough pressure provided. In this research, Cell deformation is described using liquid drop model with a constant surface tension. Different methods are proposed for determining the surface tension of individual cells and tissues such as micropipette aspiration technique [31, 32] and surface tensiometry method.[33] Based on the type of cancer, the surface tension of cancer cells varies in a wide range. For example, breast cancer cells represent soft type of cancer cells with the surface tension in the range of $0.001\ mN.m^{-1}$.[34] On the other hand, brain and cervical cancer cells are among the stiffest type of cancer cells with the surface tension of at least 1000 times greater than the breast cancer cells.[33, 35] In this study $\sigma = 27\text{mN}.\text{m}^{-1}$ is selected to represent a stiff type of cancer cells. This value is slightly greater than the surface tension of typical brain tumors. In addition, the diameter of the cancer cell is fixed at $16\mu m$ for the whole study. [36, 37] Figure 1 shows the basic design geometry of the proposed conical-shaped microfilter. As it can be seen, the device is mainly composed of three parts: entrance chamber, filtering channel (for convenience it is called microfilter) and exit chamber. Among these parts, filtering channel geometry parameters affect the efficiency of system significantly. The key parameters are: filter pore diameter ($D$) which is defined as the narrowest portion of filtering channel, pore half angle ($\theta$) and the length of the filter ($L$).

Geometry parameters for the baseline design study are listed in Table I. It is worth to mention that filter pore size diameter ($D$) has been chosen based on existing literature on effective filtering of CTCs.[16, 38, 39]

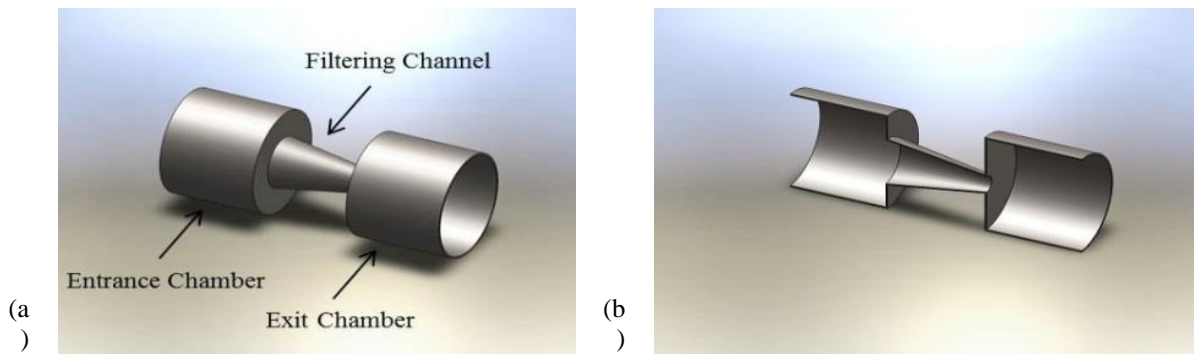

FIG. 1. (a) The Geometry of the conical-shaped microfilter device (b) A cutaway view of the device

TABLE I. Geometry parameters dimensions for the baseline design study

| Geometry Parameter | Dimensions |
| --- | --- |
| CTC diameter ($D_{CTC}(D_0)$) | 16 $\mu m$ |
| Filter pore diameter (D) | 6 $\mu m$ |
| Pore half angle (θ) | 10° |
| Filter Entrance Diameter ($D_{entrance}$) | 14.82 $\mu m$ |
| Length of the filter (L) | 25 $\mu m$ |
| Length of the entrance and exit chambers ($L_{ch}$) | 25 $\mu m$ |
| Entrance and exit chambers diameter ($D_{ch}$) | 30 $\mu m$ |

Moreover, square-wave pressure is employed in the case of studying the effects of microfilter geometry asymmetry on unclogging the system.

## III. THEORITICAL BACKGROUND

For unclogged flow of blood sample through the device, the total inlet pressure $P_{total}$ applied on the device should overcome two main pressure drop components: pressure drop due to flow of the medium and pressure drop due to surface tension of the cells when they are in contact with the wall of microfilter:

$$\Delta P_{total} = \Delta P_{flow} + \Delta P_{surface\ tension}. \tag{1}$$

### A. Pressure drop due to flow of the medium ($\Delta P_{flow}$)

The pressure drop due to flow of medium mainly consists of the pressure drop due to viscosity ($\Delta P_{hyd}$) and the pressure drop due to contraction and expansion at microfilter ($\Delta P_{contraction-expansion}$):

$$\Delta P_{flow} = \Delta P_{hyd} + \Delta P_{contraction-expansion}, \tag{2}$$

Viscous dissipation of mechanical energy of the fluid due to internal friction results in a pressure drop in the flow direction. This pressure loss can be calculated using Hagen-Poiseuille law [40] which relates the viscous pressure drop and flow rate of the flow by introducing hydraulic resistance factor. The Hagen-Poiseuille law is as follow:

$$\Delta P_{hyd} = R_{hyd} \times Q_v, \qquad (3)$$

where, $\Delta P_{hyd}$ is the viscous pressure drop of the conical channel, $R_{hyd}$ is the hydraulic resistance, and $Q_v$ is the volume flow rate. Akbari et al [41] developed a general model for predicting low Reynolds number flow pressure drop in non-uniform microchannels. Based on their model, the hydraulic resistance is calculated from the following relationship:

$$R_{hyd} = 16\pi^2 \mu \int_{x1}^{x2} \frac{I_P^*}{A(x)^2} dx, \qquad (4)$$

where $\mu$ is the viscosity of the fluid, $A(x)$ is the cross sectional area, and $I_P^* = \frac{I_p}{A^2}$ with $I_p = \int (y^2 + z^2)dA$ is called the specific polar moment of cross-sectional inertia. Based on the geometry of conical channel, $I_P^* = \frac{2}{\pi}$ and $A(x)$ can be calculated as follows:

$$A(x) = \pi r(x)^2 = \pi(Ltan(\theta) + R_a - xtan(\theta)). \qquad (5)$$

Using equations $(4-5)$, the hydraulic resistance over the length of the channel can be determined by the following formula:

$$R_{hyd} = \frac{8\mu}{3\pi \tan(\theta)} \left(\frac{1}{R_a^3} - \frac{1}{(R_a+Ltan(\theta))^3}\right). \qquad (6)$$

In addition, $\Delta P_{contraction-expansion}$ is caused by sudden contraction at the inlet of the filtering channel and sudden expansion at the outlet of filtering channel which can be calculated using the following relation:

$$\Delta P_{contraction-expansion} = K_1 \frac{\rho U_1^2}{2} + K_2 \frac{\rho U_2^2}{2}. \qquad (7)$$

where $U_1$ and $U_2$ are the flow velocity at the inlet and outlet of the filtering channel, $K_1$ is the constriction coefficient and $K_2$ is the expansion coefficient. These coefficients are set to be 0.5 and 1 respectively. It is worth to mention that pressure drop in the entrance and outlet chambers are not considered since they are negligible in comparison to two other types.

**B. Pressure drop due to surface tension ($P_{sur}$)**

Calculating the maximum pressure due to surface tension which is also referred to threshold pressure depends on whether cell is going in the forward direction (Figure 2(a)) or reverse (backward) direction (Figure 2(b)) in an alternating pressure field. Figure 2 shows the dissection of cell going through the filtering channel at critical moment at two possible directions.

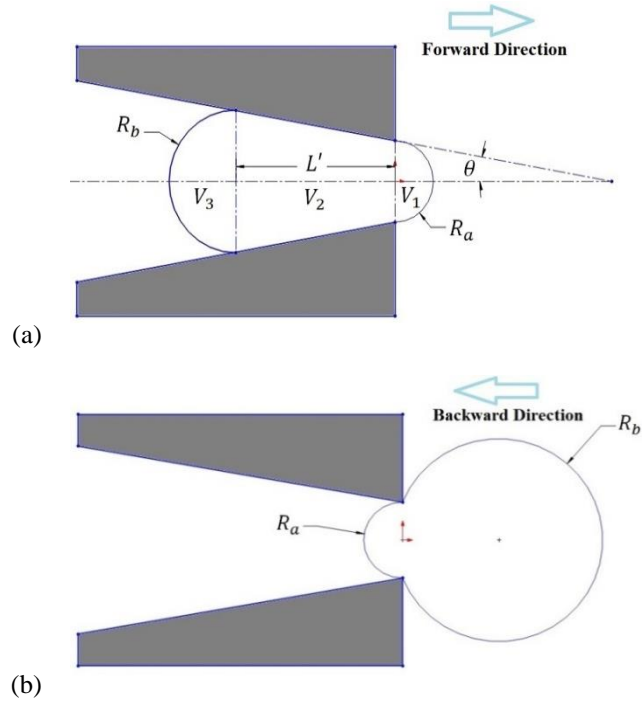

(a)

(b)

FIG. 2. 2D Cross section of microfilter at critical moment (pressure), (a) Forward passing direction (b) Backward passing direction

For both cases, the threshold pressure can be calculated using Laplace law, which relates the pressure difference between inside and outside of the cell to the cortical tension of the cell membrane by radius of curvature of the cell in the critical position.

$$P_{threshold} = 2\sigma \left(\frac{1}{R_a} - \frac{1}{R_b}\right), \tag{8}$$

where $\sigma$ is the surface tension of the cell membrane and $R_a$ and $R_b$ are the curvature radius of the leading and trailing edge respectively.

In both cases, threshold pressure occurs when the length of extension of the drop ($R_a$) out of filtering channel (forward direction) or into the filtering channel (backward direction) equals the radius of the filter pore radius ($D/2$). Therefore by knowing the surface tension of the cell, the only unknown parameter in equation (8) is $R_b$ which is different for each case.

*1. Forward threshold pressure*

In this case, $R_b$ is determined by using volume conservation law. If $V_0$ is considered as the volume of the cell with the radius of $R_0$ before squeezing into the microfilter, by modeling the geometry of the cell at the critical point of deformation in 3 parts (Figure 2(a)), volume conservation of the cell results in:

$$V_0 = V_1 + V_2 + V_3. \tag{9}$$

By considering the geometry of the microfilter, $V_1$ and $V_3$ are modeled as a half sphere of the radius of $R_a$ and $R_b$ respectively. $V_2$ can also be determined by calculating the volume of a truncated cone.

$$V_2 = \frac{\pi}{3\tan(\theta)}(R_b^3 - R_a^3). \tag{10}$$

Substituting the corresponding volumes into the equation (9) and solving for $R_b$ results in

$$R_b = \sqrt[3]{\frac{4R_0^3 - \left(2 - \frac{1}{\tan(\theta)}\right)R_a^3}{2 + \frac{1}{\tan(\theta)}}}, \tag{11}$$

*2. Backward threshold pressure*

For calculating backward threshold pressure, it is assumed that there is a negligible change in the shape of the cell in the critical moment in comparison with the initial shape of the cell before squeezing through the microfilter. Thus, the curvature radius of trailing edge will be equal to the initial radius of the cell before squeezing through the channel ($R_b = R_0$).

It should be noted that the forward and backward threshold pressure are derived based on this assumption that the cell is totally in contact with the microfilter wall which is an acceptable assumption at low Reynolds numbers.

## IV. NUMERICAL METHOD

Commercial CFD package ANSYS Fluent [42] is used for transient modeling of the CTCs passing events through the microfilter. Because the dimensions are small and the flow is multiphase, the double-precision option is enabled in order to have high accuracy. Numerical simulation is based on liquid drop model for the cell with given surface tension. Volume of fluid (VOF) method is employed for transient tracking of the interface between primary (medium) and secondary (cell) phases. The VOF method for multiphase flows can support the liquid drop model and precisely predict the pressure drop due to surface tension by Laplace law. Patching is used to define the initial position of the cell in the simulation domain. In the case of applying square-wave oscillatory inlet pressure, ANSYS user defined function (UDF) is employed to generate the transient pressure profile.

In this problem, the primary and secondary phases are surrounding medium and cell respectively. The VOF method employs volume fraction continuity equation for interface tracking:

$$\frac{1}{\rho_{cell}}\left[\frac{\partial}{\partial t}(\alpha_{cell}\rho_{cell}) + \nabla \cdot (\alpha_{cell}\rho_{cell}\overrightarrow{v_{cell}})\right] = S_{\alpha_{cell}} + (\dot{m}_{medium-cell} - \dot{m}_{cell-medium})\right], \qquad (12)$$

where $\alpha_{cell}$ is the volume fraction of the cell ($\alpha_{cell} = 0$ in medium, $\alpha_{cell} = 1$, $0 < \alpha_{cell} < 1$ at interface), $\dot{m}_{cell-medium}$ is the mass transfer from secondary phase (cell) to the primary phase (medium) and $\dot{m}_{medium-cell}$ is the mass transfer from the primary phase to secondary phase. The source term $S_{\alpha_{cell}}$ is zero in this problem.

Then, the volume fraction for medium can be computed by the following constraint:

$$\alpha_{cell} + \alpha_{medium} = 1. \qquad (13)$$

Explicit time discretization method is employed for applying VOF method because of its higher accuracy for transient tracking of the cell in comparison with implicit scheme. In order to have more stable results the courant number criteria is set within $0.25 - 0.5$.

A single momentum equation is solved throughout the domain:

$$\frac{\partial}{\partial t}(\rho\vec{v}) + \nabla \cdot (\rho\vec{v}\vec{v}) = -\nabla p + \nabla \cdot [\mu(\nabla\vec{v} + \nabla\vec{v}^T)] + \rho g + \vec{F}. \qquad (14)$$

The general properties in the flow equations can be calculated using volume fraction weighted average method. For instance, the density at an arbitrary node can be written as:

$$\rho = \alpha_{cell}\rho_{cell} + (1 - \alpha_{cell})\rho_{medium}. \qquad (15)$$

Using the divergence theorem, the surface tension force is given by:

$$F_{vol} = \sigma \frac{\rho k \nabla \alpha}{\frac{1}{2}(\rho_{cell}+\rho_{medium})}, \qquad (16)$$

where the curvature, $k = \nabla \cdot (\frac{\vec{n}}{|\vec{n}|})$, is defined in terms of the surface normal $\vec{n}$.

Moreover, wall adhesion boundary condition is included in the model for adjustment of the curvature of the surface near the wall. Wall adhesion is the surface force between the cell membrane and microfilter rigid walls when they are in contact. [32] ANSYS Fluent wall adhesion model uses the model proposed by Brackbill et al..[43] If $\theta_w$, $\hat{n}_w$ and $\hat{t}_w$ are contact angle at the wall and the unit vectors normal and tangential to the wall respectively, we have:

$$\hat{n} = \hat{n}_w \cos(\theta_w) + \hat{t}_w \sin(\theta_w), \qquad (17)$$

In this study, $\theta_w$ is selected as 180° for phase interaction. To speed up the solution process, non-iterative time advancement (NITA) scheme is used, employing fractional step method.

With respect to the geometry of the device, a 2D axisymmetric geometry is employed in order to reduce the cost of calculation. The mesh is built using ANSYS meshing. The mesh is refined along the walls of the microfilter and symmetry axis in order to increase the accuracy. The total number of element used for simulation is around 15,000 which are mainly quadratic elements. Figure 3 shows the mesh structure of the geometry.

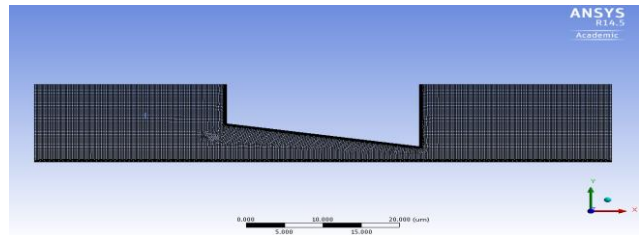

FIG. 3. 2D axisymmetric mesh structure of the geometry

## V. RESULT AND DISCUSSION

### A. Numerical method validation

As discussed in previous sections, we model CTC as a liquid droplet with given surface tension passing through the microfilter using VOF model in Fluent. To validate the numerical simulation, we compared our results with the experimental work done by Guo et al..[30] They studied the deformation of single cells passing through rectangular-shaped microfilters and measured the required pressure to squeeze the cell through the channel called threshold pressure. Based on the results and using liquid drop model, they calculated the surface tension of the mouse lymphoma cell (MLC). For comparison purpose, $0.75\ mN.m^{-1}$ is set for the surface tension of the cell based on their results and compared the required threshold pressures obtained from numerical simulation and measured from experiments for different microfilter pore sizes. The results, as shown in Figure 4, indicate that the numerical threshold pressures match quite well with experimental threshold pressures. The values obtained from numerical simulations are only slightly smaller than that of experiments. This is mainly because in our method the cell is considered as a homogenous liquid and the intracellular structure which resists the deformation is not considered. Based on the result, it can be concluded that by knowing the surface tension of CTCs, we can predict the pressure of the cell filtration process using our liquid-drop based numerical model. It should be noted that the selected value of $0.75\ mN.m^{-1}$ is just used in this section for comparing the numerical simulation with the experiment. In following sections, $\sigma = 27\text{mN}.\text{m}^{-1}$ is fixed for CTC modeling since the main focus of the present research is on the application of such a device on human CTC separation.

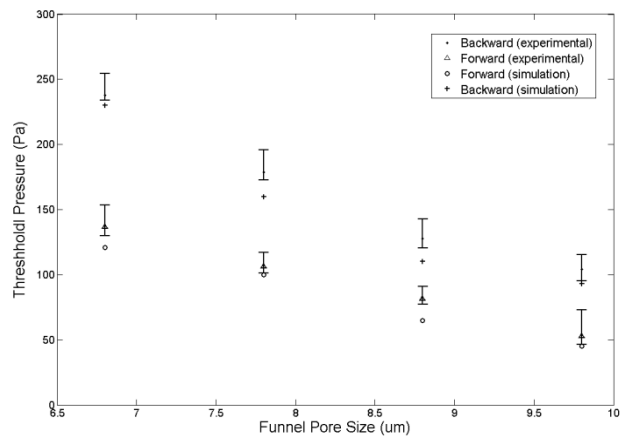

FIG. 4. Comparison of experimental and numerical threshold pressures for mouse lymphoma cell (MLC) passing event through channels with different funnel pore size

**B. Pressure-deformability analysis**

The first step in designing a device and optimizing its performance is to completely understand the fundamental concepts behind the working principle of the system. In deformability-based CTC separation devices, squeezing a CTC through the microfilter is the most important part of separation process. Thus, understanding thoroughly how a cancer cell behaves in this process is of great importance. In this section, we study the whole process of a cancer cell passing through a conical-shaped microfilter by obtaining the pressure signature of the system and the corresponding cell deformation throughout the process. Based on two possible moving directions of the cell, forward or backward direction, the pressure-deformability behavior varies due to geometry asymmetry. In both cases, pressure variation can be explained by the deformation of the cell. It should be noted that the pressure signature refers to the time history of the total inlet pressure of the system when a cell is getting squeezed through the filter with a constant flow rate, which for convenience is called total pressure.

*1. Forward direction*

The pressure-deformability behavior of the cell passing forward through the microfilter is shown in Figure 5(a). Before the cell is going through the microfilter, the total pressure is relatively constant and is mainly due to viscous pressure (stage (a1)). As the cell starts to squeeze through the microfilter, the total pressure increases due to increase in surface tension pressure. The maximum required total pressure occurs when the length of extension of the cancer cell out of filtering channel equals the radius of the pore (stage (a2)). This maximum pressure is called critical pressure. After this stage, the radius of the curvature of the cell at leading edge starts to increase as the cell is going out of the microfilter and consequently the radius of curvature of trailing edge starts to decrease. As a result, there is a sudden decrease in total pressure after stage (a2). At stage (a3), the total pressure equals the initial viscous pressure since the surface tension resistance becomes zero. This is mainly because the radius of curvature of leading and trailing edges are approximately equal at this point. As the cell going further out of the microfilter, the curvature radius of leading edge becomes larger than trailing edge. Hence, based on Laplace law, the surface tension force is not a barrier for the flow any more, but helps to pull the cell out of the microfilter in the direction of the flow. Thus, in a constant flow process, such a phenomenon results in a negative total pressure when the curvature radius of trailing edge becomes large enough. After the cell completely passed the microfilter, there is a local maximum pressure due to the bouncing of rear part of the cell which is mainly because of fluid inertia. This phenomenon is more obvious at higher flow rate.

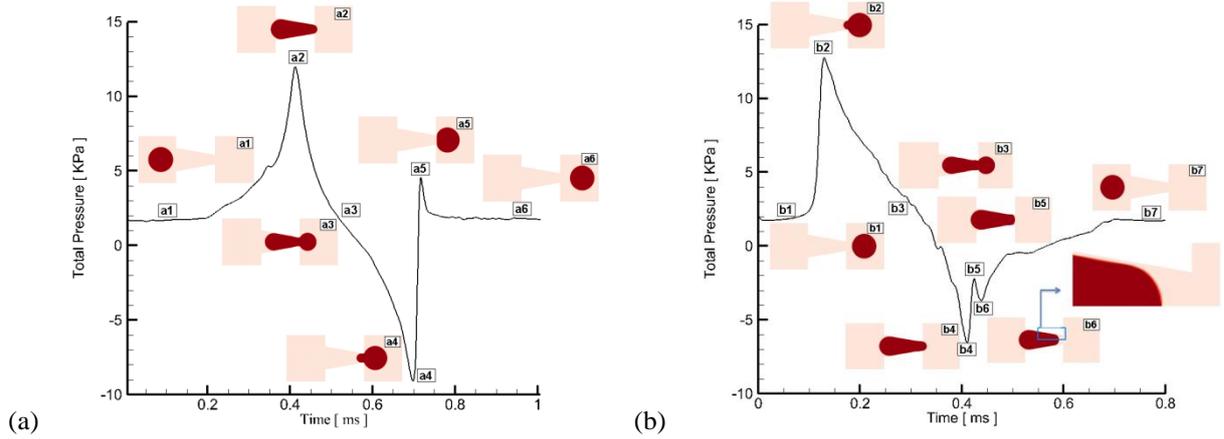

FIG. 5 (a) Pressure-deformability behaviour of the cancer cell moving through the microfilter in forward direction: (a1) initial state before entering the microfilter (a2) maximum pressure (a3) equilibrium state (a4) minimum pressure (a5) local maximum pressure due to rear part bouncing (a6) final state after exiting the microfilter (b) Pressure-deformability behaviour of the cancer cell moving through the microfilter in backward direction (b1) initial state before entering the microfilter in backward direction (b2) maximum pressure (b3) equilibrium state (b4) minimum pressure (b5) local maximum pressure (b6) local minimum pressure (b7) final state after exiting the microfilter

*2. Backward direction*

The cell behavior in backward moving direction is shown in Figure 5(b). In backward direction, maximum pressure (critical pressure) occurs when the radius of the cell into the microfilter equals the radius of the microfilter pore (stage (b2)). By comparing the two critical pressures in backward and forward directions, it is observed that the backward critical pressure is higher than the forward critical pressure. After this stage the total pressure starts to decrease due to increase in radius of leading edge and decrease in radius of trailing edge. Like in forward direction, there is an equilibrium state at stage (b3) due to equality of radius of curvature of leading and trailing edge. As it can be seen in Figure 5(b), there is a decrease in total pressure between stages (b5) and (b6). This is mainly because of the detachment of the cancer cell from the microfilter walls. As the cell detaches from the wall (stage (b5)), the radius of curvature of trailing edge starts to decrease until it reaches its minimum (stage (b6)). The decrease in trailing edge radius causes a temporary decrease in total pressure. In addition, in backward direction there is no significant pressure perturbation due to cell bouncing (stage (a5) in forward direction) since the cell turns back gradually to its initial shape with the aid of mirofilter conical shape.

**C. Critical pressure vs. system throughput**

A successful cell passing event through the microfilter is mainly dependent on the flow ability to overcome total pressure resistance consists of viscous resistance and the resistance caused by surface tension. Throughout the forward process, the critical pressure occurs when the length of extension of the drop out of the filtering channel ($R_a$) equals the radius of the filter pore radius ($D/2$). Since most of the CTC separation devices utilize pressure driven flow for separation process, having accurate information on required inlet total pressure to overcome critical pressure has gained importance; for a successful filtering process this applied pressure should be kept below the critical pressure of CTCs and above the critical pressure of white blood cells (WBCs). The other important design criterion of CTC separation devices is system throughput. Obviously higher throughput is preferred in clinical applications. Volume flow rate of the device is a clear indicator of the system throughput. In this section we studied the effect of volume flow rate on pressure components when the cell is going in forward direction (Figure 6).

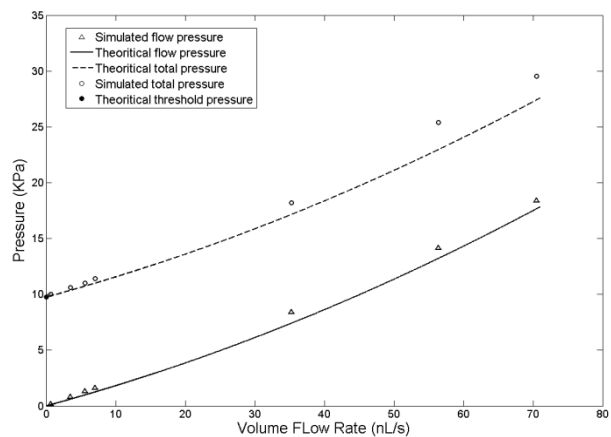

FIG. 6. The effect of volume flow rate on pressure components and total pressure of the system based on numerical simulation and theoretical formulas

The results indicate that the pressure components of the system are strongly dependent on the volume flow rate and as a results system throughput. Moreover, the simulation results match well with theoretical pressures at low volume flow rates. The results show that total and viscous pressure increase with increasing in volume flow rate at roughly the same speed. Given this result, it can be concluded that pressure resistance due to surface tension is not dependent on volume flow rate and is relatively constant. Therefore the increase in total pressure is mainly due to increase in viscous pressure as a result of changing in flow velocity. As it can be seen from Figure 6, at higher volume flow rates, there is a deviation between the total inlet pressure calculated through theoretical formulas and the one gained by numerical simulation; this is mainly because of limiting assumptions made to derive the theoretical formulas which should be considered at higher flow rates. For example, $\Delta P_{hyd}$ is calculated by neglecting the presence of the cell inside the pore. Such an assumption works well in low flow rates since $\Delta P_{hyd}$ is proportional to the flow rate but in higher flow rates it causes significant deviation from numerical simulation results.

Based on the results, in order to increase the system throughput (increasing volume flow rate), the critical pressure also increases which may change the cytoskeleton structure and cause cellular damages or ruptures especially in CTC separation process due to fragility of cancer cells. This is one of the operating limitations of the filtering process which has to be considered and a trade-off should be done between increasing system throughput and avoiding permanent cellular damages since one of the main performance criteria of a CTC separation device is its ability to offer researcher viable cells in order to have a wider range of post processing analysis.[8, 10, 44] Due to comparison purpose, we fix the volume flow rate at $7 nL/s$ for the following studies.

**D. System unclogging**

One of the key challenges in CTC separation devices is clogging which limits the overall selectivity of separation process. One of the approaches to solve clogging is to apply a periodically reversed pressure. The effectiveness of this approach is limited in constant cross-sectional microfilters since it significantly reduces the net displacement of cells in forward direction and accordingly system throughput especially at low-Reynolds flows. Based on pressure-deformability results, in conical-shaped microfilters the critical pressure in backward direction is higher than the corresponding forward one in which the difference is dependent on the microfilter geometry. This feature enables us to unclog these devices by applying a periodically reversed pressure. This applied pressure should have the amplitude higher than the forward critical pressure in order for the cell to squeeze through the channels and lower than backward critical pressure in the case of avoiding undoing the separation process. In this regard, choosing suitable microfilter geometry parameters and system applied pressure enables us to unclog the system in an efficient way. In this section we study the effect of pore half angle as one of the microfilter geometry parameters and system applied pressure on the efficiency of the unclogging process.

*1. Effect of pore half angle on unclogging the system*

In order to quantitatively evaluate this feature we define pressure ratio ($PR$) as the ratio of backward critical pressure to forward critical pressure.

High pressure ratios enable us to select the suitable applied pressure in a wider range. Thus, finding the optimum $PR$ is required in order to improve the device performance. In this section we study the effect of pore half angle on backward and forward critical pressure which leads to calculating $PR$. In order to perform this analysis we fixed other geometric parameters and operating conditions.

Figure 7 shows the variation of forward and backward critical pressure and the resulting pressure ratio with respect to pore half angle (Figure 7(a)) and its corresponding cell deformation at critical moment (Figure 7(b)). As it can be seen, the critical pressure mainly depends on how cell has been deformed at critical moment.

*a. Critical forward pressure vs. pore half angle*

At low angles (starting with 2°), the critical moment happens when the large portion of the cell is still out of microfilter entrance and cell has contact with entrance wall of the microfilter. As a result, large radius of curvature of trailing edge causes large critical pressure. By increasing the angle, the portion of the cell out of the microfilter entrance and consequently the radius of the trailing edge become smaller which results in a decrease in critical pressure. This process continues until pore half angle of 7° where the cell has no contact with the entrance wall of microfilter anymore at critical moment. As the half angle of pore is further increased, the leading edge of cell at critical moment goes further through the microfilter and becomes smaller following the geometry of the channel, and accordingly results in a further decrease in forward critical pressure but with different slope. At microfilter angle of 12°, the entrance of microfilter is larger than the diameter of the cell, and the cell passes through the microfilter without any contact with the microfilter entrance wall. Such a large pore half angle also causes to have a layer of water between the microfilter inner wall and the cell. As a result, a sudden decrease occurs in forward critical pressure. By increasing the pore half angle further than 12°, the existing layer of water between the cell and the inner walls of microfilter becomes smaller. This causes an increase in leading edge radius and also trailing edge radius with different slope. Thus, there is an increase in forward critical pressure. At higher angles (higher than 15°), there is no significant variation due to pore half angle because the cell deformation at critical moments doesn't change very significantly.

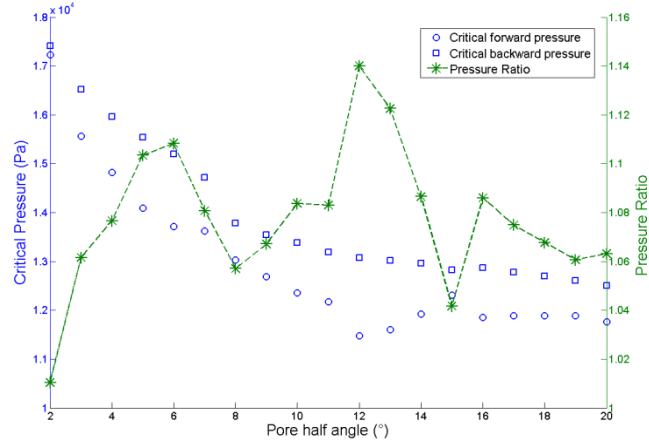

(a)

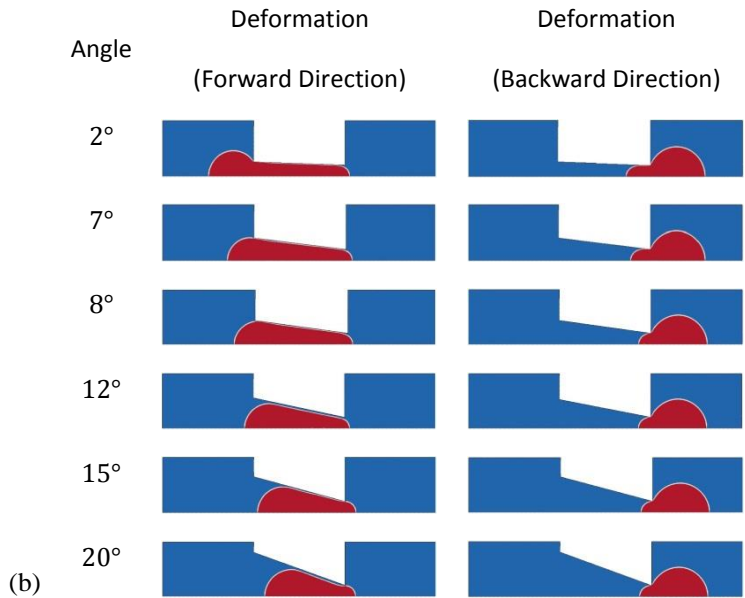

(b)

FIG. 7. (a) Critical Pressures (forward and backward directions) and the corresponding pressure ratios of the system with respect to pore half angle (b) Corresponding deformation of the passing cell at critical moment for key pore half angles

*b. Critical backward pressure vs. pore half angle*

Like the forward direction, the behavior of cell and consequently the backward critical pressure mainly depends on the geometry of the cell at critical moment. As the pore half angle is increasing, the portion of the cell into the microfilter at critical moment becomes smaller since the cell loses its contact with inner wall of the microfilter. Accordingly, the radius of leading edge increases which causes a decrease in backward critical pressure. At pore half angle of 8° the leading edge of the cell totally loses its contact with inner wall of the microfilter. Therefore, there is a sudden decrease in both the portion of the cell into the channel and critical pressure. By increasing the angle of the pore, decreasing in critical pressure continues but with a slower slope since the changes in leading and trailing edge radius is not significant.

Based on these results we can derive the variation of *PR* with respect to pore half angle. The results are shown in Figure 7(a). As it can be seen, the pressure ratio does not follow a specific trend; this is mainly because the cell deformation at critical moment, which has a direct relation to the critical pressure, behaves differently at each direction and angle. According to the results, the maximum *PR* occurs at angle of 12. Although there exist another local maximum at 6°, the microfilter angle of 12° is more preferable since in this case the radius of microfilter entrance is bigger than the cell radius which avoid the long contact of the cell with the microfilter walls.

*2. System performance under unclogging applied pressure*

As discussed in previous sections, pressure asymmetry behavior of conical shaped microfilters enables us to unclog the system by applying a periodically reversed-pressure. But the type of applied pressure and its characteristics affect system performance significantly. In this section, square wave profile is selected for the applied pressure and the effects of profile duty cycle and frequency as two important square wave parameters on system performance are investigated. For this purpose, the cancer cell displacement as an indicator of system throughput is measured with respect to flow time for different applied pressures. It should be noted that, the square wave amplitude for all of selected applied pressure is fixed slightly higher than critical forward pressure and much lower than critical backward pressure in order to avoid undoing the filtration process. Generally, by applying forward pressure on the system, the cell is going through the microfilters since the pressure is higher than the required critical forward pressure, while in reversed regime, the cell stops moving when it reaches the microfilter pore.

*a. Effect of square wave duty cycle on system performance*

Square wave duty cycle is defined as the ratio of forward to reversed time steps in a period. Four different duty cycles with constant time period of $1ms$ are selected. Figure 8 shows cancer cell displacement with respect to flow time (Figure 8 (a)) and the position of the cancer cells at the last time step (Figure 8 (b)-(e)) for different applied pressures. As expected, with increasing the duty cycle, the net displacement of the cell and as a result system throughout increase but the possibility of unclogging decreases due to decrease in the applied time of reversed pressure. On the other hand, in lower duty cycles the cell net displacement decreases and even the cell may get stuck between two consecutive microfilter and move periodically between them like in duty cycle of 1. As a result, a tradeoff should be done between the cell net displacements and unclogging of the system.

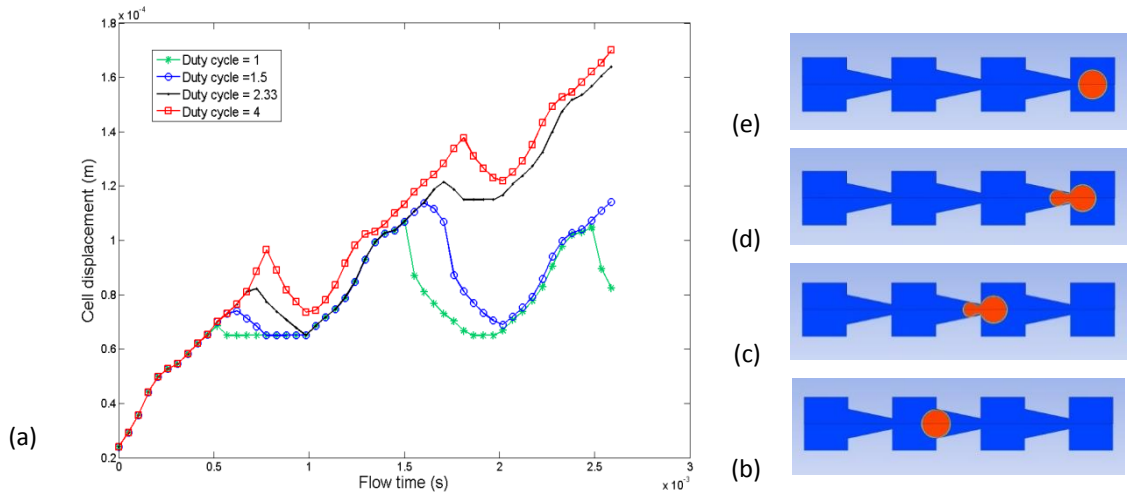

FIG. 8. Effect of square wave duty cycle on system performance (a) Cell displacement vs. flow time (b) corresponding cell position at final time step for duty cycle of 1 (c) duty cycle of 1.5 (d) duty cycle of 2.33 (e) duty cycle of 4

*b. Effect of square wave frequency on system performance*

Square wave frequency is another determinant factor that greatly affects the system behavior. In this section, square wave profile with duty cycle of 4 and frequency of $\omega$ is selected as the most efficient profile for increasing the system throughput based on the results of previous section and studied the effect of profile frequency on cell displacement. Thus, three different profiles with the same duty cycles and different frequencies are selected, as shown in Figure 9(a). The results, shown in Figure 9(b), illustrates that decreasing the frequency results in further increase in net cell displacement and accordingly system throughput while in higher frequencies the cell net displacement reduces and even it can be zero.

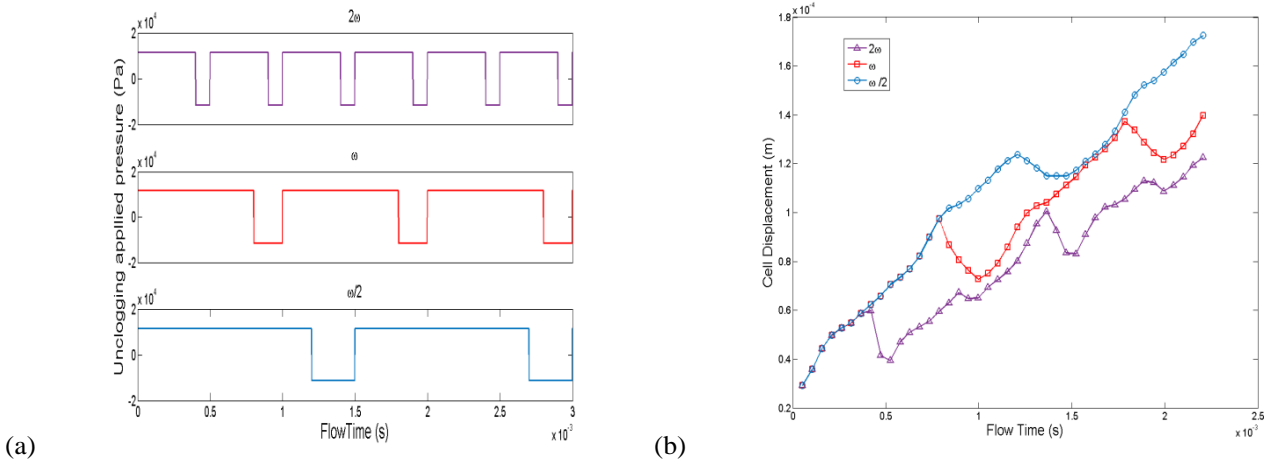

(a)                                                                         (b)

FIG. 9. (a) Square wave profiles with different frequencies (b) Cell displacement vs. flow time for different frequencies

## VI. MODEL LIMITATIONS AND FUTURE WORKS

In the present research, the liquid drop model is used as the mechanical model of living cells. Although such a model is very powerful in predicting the critical pressure of the system as one of the main determinant design criteria of deformability-based CTC separation microfilters, like other proposed mechanical models it cannot predict the behavior of living cells in all aspects. The tendency of cells to stick on surfaces and pores once they are trapped is one of the major phenomena that cannot be predicted by the liquid drop model. Overcoming such a phenomenon requires more complex mechanical models, such as coupling the liquid drop model with proposed models for adhesion of living cells to surfaces. [45, 46] Furthermore, the liquid drop model results in a faster cell passing event through microfilters than the real process due to neglecting the intercellular structure of living cells. This issue is more obvious in frequency analysis of the system in which it results in higher working frequencies. Thus, the liquid drop model is not suitable for quantitative frequency analysis. It should be noted that such a limitation does not affect the frequency analysis results in the present research since the main focus is on the qualitative comparison of different applied pressure profiles.

Future works will involve more accurate mechanical models of living cells in terms of the ability to predict more aspects of living cells. In addition, although simulating the passing event of only one CTC can be a good assumption, due to the rarity of such cells in the bloodstream of people with cancer (one CTC in a billion normal blood cells[47]), the effects of multiple pores and cells should be considered in order to perform more accurate evaluation on the overall throughput and clogging of the system.

## VII. CONCLUSION

Non-uniform cross-sectional microfilters have the advantage of pressure asymmetry which makes unclogging process of CTC separation devices more efficient than uniform cross-sectional microfilters. We adopted numerical simulation as a useful tool to study conical-shaped microfilters. In first step, we validated our model with an experimental work done on deformation of individual cells going through a rectangular channel. The numerical results compare favorably with experimental data indicating that our numerical model can predict well the critical pressure of the process. Then, we analyzed pressure-deformability behavior of the cancer cell going through the microfilter via two possible directions (forward and backward) in order to better understand the squeezing process; based on this analysis we identified the critical moments and their corresponding pressures and deformations in both directions. More importantly, we found out that the backward critical pressure is higher than the forward critical pressure which results in a pressure asymmetry behavior in the conical-shaped microfilter. This feature gives such devices the ability of unclogging by applying a suitable inverse pressure. Next, we studied the effect of system throughput on the required inlet pressure. We observed that by increasing system throughput the required inlet pressure increases as well. This leads to a tradeoff procedure between the system throughput and cellular damage considerations. In the next step, two important aspects of unclogging issue were discussed. First, we studied the effect of pore half angle on pressure asymmetry by defining pressure ratio, and determined the optimum pore half angle for maximizing the pressure ratio through parametric analysis. Second, we investigated the effect of duty cycle and frequency of the square-wave applied pressure on system performance. The numerical results indicate that while lower duty cycles offer higher possibility of unclogging, they also reduce the throughput of the system which is undesirable. As a result, a tradeoff should be made in order to choose the optimum duty cycle based on system requirements. Furthermore, lower frequency of applied pressure results in a higher cell displacement and accordingly higher system throughput. Finally, limitations of the model were discussed and possible future works were presented.

**REFERENCES**

[1] Ma, X., and Yu, H., "Global burden of cancer," The Yale journal of biology and medicine, 79(3-4), pp. 85-94 (2006).

[2] Siegel, R., Naishadham, D., and Jemal, A., "Cancer statistics, 2013," CA: A Cancer Journal for Clinicians, 63(1), pp. 11-30 (2013).

[3] Popat, K., McQueen, K., and Feeley, T. W., "The global burden of cancer," Best Practice & Research Clinical Anaesthesiology, 27(4), pp. 399-408 (2013).

[4] Liberko, M., Kolostova, K., and Bobek, V., "Essentials of circulating tumor cells for clinical research and practice," Critical Reviews in Oncology/Hematology, 88(2), pp. 338-356 (2013).

[5] Harouaka, R., Kang, Z., Zheng, S.-Y., and Cao, L., "Circulating tumor cells: Advances in isolation and analysis, and challenges for clinical applications," Pharmacology & Therapeutics, 141(2), pp. 209-221 (2014).

[6] Jin, C., McFaul, S. M., Duffy, S. P., Deng, X., Tavassoli, P., Black, P. C., and Ma, H., "Technologies for label-free separation of circulating tumor cells: from historical foundations to recent developments," Lab on a Chip, 14(1), pp. 32-44 (2014).

[7] Allan, A. L., and Keeney, M., "Circulating tumor cell analysis: technical and statistical considerations for application to the clinic," Journal of oncology, 2010 (2009).

[8] Gossett, D., Weaver, W., Mach, A., Hur, S., Tse, H., Lee, W., Amini, H., and Di Carlo, D., "Label-free cell separation and sorting in microfluidic systems," Anal Bioanal Chem, 397(8), pp. 3249-3267 (2010).

[9] Li, P., Stratton, Z. S., Dao, M., Ritz, J., and Huang, T. J., "Probing circulating tumor cells in microfluidics," Lab on a Chip, 13(4), pp. 602-609 (2013).

[10] Chen, Y., Li, P., Huang, P.-H., Xie, Y., Mai, J. D., Wang, L., Nguyen, N.-T., and Huang, T. J., "Rare cell isolation and analysis in microfluidics," Lab on a Chip, 14(4), pp. 626-645 (2014).

[11] Hur, S. C., Henderson-MacLennan, N. K., McCabe, E. R., and Di Carlo, D., "Deformability-based cell classification and enrichment using inertial microfluidics," Lab on a Chip, 11(5), pp. 912-920 (2011).

[12] Hochmuth, R. M., "Micropipette aspiration of living cells," Journal of Biomechanics, 33(1), pp. 15-22 (2000).

[13] Gerhardt, T., Woo, S., and Ma, H., "Chromatographic behaviour of single cells in a microchannel with dynamic geometry," Lab on a Chip, 11(16), pp. 2731-2737 (2011).

[14] McFaul, S. M., Lin, B. K., and Ma, H., "Cell separation based on size and deformability using microfluidic funnel ratchets," Lab on a chip, 12(13), pp. 2369-2376 (2012).

[15] Zheng, S., Lin, H., Liu, J.-Q., Balic, M., Datar, R., Cote, R. J., and Tai, Y.-C., "Membrane microfilter device for selective capture, electrolysis and genomic analysis of human circulating tumor cells," Journal of Chromatography A, 1162(2), pp. 154-161 (2007).

[16] Zhang, Z., Xu, J., Hong, B., and Chen, X., "The effects of 3D channel geometry on CTC passing pressure - towards deformability-based cancer cell separation," Lab on a Chip, 14(14), pp. 2576-2584 (2014).

[17] Tan, S., Yobas, L., Lee, G., Ong, C., and Lim, C., "Microdevice for the isolation and enumeration of cancer cells from blood," Biomedical Microdevices, 11(4), pp. 883-892 (2009).

[18] Tang, Y., Shi, J., Li, S., Wang, L., Cayre, Y. E., and Chen, Y., "Microfluidic device with integrated microfilter of conical-shaped holes for high efficiency and high purity capture of circulating tumor cells," Sci. Rep., 4 (2014).

[19] Gross, B. C., Erkal, J. L., Lockwood, S. Y., Chen, C., and Spence, D. M., "Evaluation of 3D Printing and Its Potential Impact on Biotechnology and the Chemical Sciences," Analytical Chemistry, 86(7), pp. 3240-3253 (2014).

[20] Boey, S. K., Boal, D. H., and Discher, D. E., "Simulations of the erythrocyte cytoskeleton at large deformation. I. Microscopic models," Biophysical Journal, 75(3), pp. 1573-1583 (1998).

[21] Coughlin, M. F., and Stamenović, D., "A prestressed cable network model of the adherent cell cytoskeleton," Biophysical Journal, 84(2 I), pp. 1328-1336 (2003).

[22] Satcher Jr, R. L., and Dewey Jr, C. F., "Theoretical estimates of mechanical properties of the endothelial cell cytoskeleton," Biophysical Journal, 71(1), pp. 109-118 (1996).

[23] Stamenović, D., Fredberg, J. J., Wang, N., Butler, J. P., and Ingber, D. E., "A microstructural approach to cytoskeletal mechanics based on tensegrity," Journal of Theoretical Biology, 181(2), pp. 125-136 (1996).

[24] Stamenović, D., and Ingber, D. E., "Models of cytoskeletal mechanics of adherent cells," Biomech Model Mechanobiol, 1(1), pp. 95-108 (2002).

[25] Lim, C. T., Zhou, E. H., and Quek, S. T., "Mechanical models for living cells—a review," Journal of Biomechanics, 39(2), pp. 195-216 (2006).

[26] Yeung, A., and Evans, E., "Cortical shell-liquid core model for passive flow of liquid-like spherical cells into micropipets," Biophysical Journal, 56(1), pp. 139-149 (1989).

[27] Leong, F., Li, Q., Lim, C., and Chiam, K.-H., "Modeling cell entry into a micro-channel," Biomech Model Mechanobiol, 10(5), pp. 755-766 (2011).

[28] Zhang, Z., Chen, X., and Xu, J., "Entry effects of droplet in a micro confinement: Implications for deformation-based circulating tumor cell microfiltration," Biomicrofluidics, 9(2), p. 024108 (2015).

[29] Luo, Y. N., Chen, D. Y., Zhao, Y., Wei, C., Zhao, X. T., Yue, W. T., Long, R., Wang, J. B., and Chen, J., "A constriction channel based microfluidic system enabling continuous characterization of cellular instantaneous Young's modulus," Sensors and Actuators B: Chemical, 202(0), pp. 1183-1189 (2014).

[30] Guo, Q., McFaul, S. M., and Ma, H., "Deterministic microfluidic ratchet based on the deformation of individual cells," Physical Review E, 83(5), p. 051910 (2011).

[31] Guo, Q., Park, S., and Ma, H., "Microfluidic micropipette aspiration for measuring the deformability of single cells," Lab on a Chip - Miniaturisation for Chemistry and Biology, 12(15), pp. 2687-2695 (2012).

[32] Kothe, D. B., and Mjolsness, R. C., "RIPPLE - A new model for incompressible flows with free surfaces," AIAA Journal, 30(11), pp. 2694-2700 (1992).

[33] Winters, B. S., Shepard, S. R., and Foty, R. A., "Biophysical measurement of brain tumor cohesion," International Journal of Cancer, 114(3), pp. 371-379 (2005).

[34] Guo, H.-L., Liu, C.-X., Duan, J.-F., Jiang, Y.-Q., Han, X.-H., Li, Z.-L., Cheng, B.-Y., and Zhang, D.-Z., "Mechanical Properties of Breast Cancer Cell Membrane Studied with Optical Tweezers," Chinese Physics Letters, 21(12), p. 2543 (2004).

[35] Preetha, A., Huilgol, N., and Banerjee, R., "Interfacial properties as biophysical markers of cervical cancer," Biomedicine & Pharmacotherapy, 59(9), pp. 491-497 (2005).

[36] Hou, H. W., Warkiani, M. E., Khoo, B. L., Li, Z. R., Soo, R. A., Tan, D. S.-W., Lim, W.-T., Han, J., Bhagat, A. A. S., and Lim, C. T., "Isolation and retrieval of circulating tumor cells using centrifugal forces," Sci. Rep., 3 (2013).

[37] Meng, S., Tripathy, D., Frenkel, E. P., Shete, S., Naftalis, E. Z., Huth, J. F., Beitsch, P. D., Leitch, M., Hoover, S., Euhus, D., Haley, B., Morrison, L., Fleming, T. P., Herlyn, D., Terstappen, L. W. M. M., Fehm, T., Tucker, T. F., Lane, N., Wang, J., and Uhr, J. W., "Circulating Tumor Cells in Patients with Breast Cancer Dormancy," Clinical Cancer Research, 10(24), pp. 8152-8162 (2004).

[38] Gusenbauer, M., Cimrak, I., Bance, S., Exl, L., Reichel, F., Oezelt, H., and Schrefl, T., "A tunable cancer cell filter using magnetic beads: cellular and fluid dynamic simulations," arXiv preprint arXiv:1110.0995 (2011).

[39] Mohamed, H., Murray, M., Turner, J. N., and Caggana, M., "Isolation of tumor cells using size and deformation," Journal of Chromatography A, 1216(47), pp. 8289-8295 (2009).

[40] Bruus, H., Theoretical Microfluidics, Oxford University Press, (2008).

[41] Akbari, M., Sinton, D., and Bahrami, M., "Viscous flow in variable cross-section microchannels of arbitrary shapes," International Journal of Heat and Mass Transfer, 54(17–18), pp. 3970-3978 (2011).

[42] ANSYS Fluent 14.5.0 Documentation, ANSYS® Academic Research, Release 14.5.0, ANSYS, Inc., (2011).

[43] Brackbill, J. U., Kothe, D. B., and Zemach, C., "A continuum method for modeling surface tension," Journal of Computational Physics, 100(2), pp. 335-354 (1992).

[44] Kuo, J. S., Zhao, Y., Schiro, P. G., Ng, L., Lim, D. S. W., Shelby, J. P., and Chiu, D. T., "Deformability considerations in filtration of biological cells," Lab on a Chip, 10(7), pp. 837-842 (2010).

[45] Hammer, D. A., and Apte, S. M., "Simulation of cell rolling and adhesion on surfaces in shear flow: general results and analysis of selectin-mediated neutrophil adhesion," Biophysical Journal, 63(1), pp. 35-57 (1992).

[46] N'Dri, N. A., Shyy, W., and Tran-Son-Tay, R., "Computational Modeling of Cell Adhesion and Movement Using a Continuum-Kinetics Approach," Biophysical Journal, 85(4), pp. 2273-2286 (2003).

[47] Yu, M., Stott, S., Toner, M., Maheswaran, S., and Haber, D. A., "Circulating tumor cells: approaches to isolation and characterization," The Journal of Cell Biology, 192(3), pp. 373-382 (2011).